\documentclass[conference]{IEEEtran} 
\IEEEoverridecommandlockouts
\usepackage{cite}
\usepackage{amsmath,amssymb,amsfonts}
\usepackage{algorithmic}
\usepackage{textcomp}

\usepackage{graphicx}
\usepackage{color, soul, xcolor}  
\usepackage{colortbl}
\usepackage{booktabs} 
\usepackage{enumitem}
\usepackage{caption}
\usepackage{amsmath}
\usepackage{listings}
\usepackage{balance}
\usepackage{caption}
\usepackage[caption=false,font=footnotesize,labelformat=simple]{subfig}
\usepackage[normalem]{ulem} 
\usepackage{framed}
\definecolor{shadecolor}{gray}{0.97}

\usepackage[linewidth=1pt]{mdframed} 
\usepackage{romannum}

\newcommand{\MYROMAN}[1]{%
  \textup{\uppercase\expandafter{\romannumeral#1}}%
}

\def\BibTeX{{\rm B\kern-.05em{\sc i\kern-.025em b}\kern-.08em
    T\kern-.1667em\lower.7ex\hbox{E}\kern-.125emX}}

\begin{document}

\title{\textit{\Romannum{11} Commandments of Kubernetes Security}: A Systematization of Knowledge Related to Kubernetes Security Practices} 

\author{\IEEEauthorblockN{Md. Shazibul Islam Shamim}
\IEEEauthorblockA{\textit{Dept. of Computer Science} \\
\textit{Tennessee Technological University}\\
Cookeville, TN, USA\\
mshamim42@students.tntech.edu}
\and
\IEEEauthorblockN{Farzana Ahamed Bhuiyan}
\IEEEauthorblockA{\textit{Dept. of Computer Science} \\
\textit{Tennessee Technological University}\\
Cookeville, TN, USA \\
fbhuiyan42@students.tntech.edu}
\and
\IEEEauthorblockN{Akond Rahman}
\IEEEauthorblockA{\textit{Dept. of Computer Science} \\
\textit{Tennessee Tech. University}\\
Cookeville, TN, USA\\
arahman@tntech.edu}
}

\maketitle

\begin{abstract}

Kubernetes is an open-source software for automating management of computerized services. Organizations, such as IBM, Capital One and Adidas use Kubernetes to deploy and manage their containers, and have reported benefits related to deployment frequency. Despite reported benefits, Kubernetes deployments are susceptible to security vulnerabilities, such as those that occurred at Tesla in 2018. A systematization of Kubernetes security practices can help practitioners mitigate vulnerabilities in their Kubernetes deployments. \textit{The goal of this paper is to help practitioners in securing their Kubernetes installations through a systematization of knowledge related to Kubernetes security practices.} We systematize knowledge by applying qualitative analysis on 104 Internet artifacts. We identify 11 security practices that include (i) implementation of role-based access control (RBAC) authorization to provide least privilege, (ii) applying security patches to keep Kubernetes updated, and (iii) implementing pod and network specific security policies.  \\

{\color{red} The paper has recently been accepted at the IEEE SecDev 2020 Conference on June 26, 2020. }

\end{abstract}



\section{Introduction}
\label{intro}

Kubernetes is an open-source software for automating management of computerized services, such as containers~\cite{miles:book2020:kubernetes}. Practitioners use Kubernetes because it reduces repetitive manual processes involved in container deployment and management. Kubernetes is considered one of the most popular open-source container orchestration tools and it is used in organizations such as Adidas, Nokia, Spotify, and the U.S. Department of Defense (DoD)~\cite{k8s:case:studies, cncf:case:studies}. Benefits of Kubernetes usage have been documented: for example usage of Kubernetes in the U.S. DoD resulted in reducing an eight month software deployment effort down to one week~\cite{cncf:case:studies}. For Adidas, the load time for an e-commerce website was reduced by half, and release frequency increased from once every 4$\mathtt{\sim}$6 weeks to 3$\mathtt{\sim}$4 times a day~\cite{k8s:case:studies}. 

Despite reported benefits, Kubernetes users have reported their concerns related to Kubernetes security. The Cloud Native Computing Foundation conducted a survey with 1,337 practitioners and reported 40\% of the survey participants to be concerned with Kubernetes security~\cite{cncf:survey}. Anecdotal evidence supports practitioner-reported concerns related to Kubernetes security. For example, in 2018, malicious users gained access to Tesla's Amazon Web Services (AWS) resources using an insecure Kubernetes console~\cite{tesla:k8s:attack}.   

Systematizing available knowledge regarding Kubernetes security practices could support practitioners in securing their Kubernetes installations.  Such systematization of knowledge can be beneficial for practitioners who (i) want to understand what activities need to be executed to secure Kubernetes components and (ii) can use the derived list of practices as a benchmark to compare their state of security practices.  

Systematization of knowledge can be conducted by analyzing Internet artifacts, such as blog posts and video presentations. Instead of academic forums, such as research conferences, practitioners often report what practices they use in Internet artifacts~\cite{garousi2018smells:blog, glass2006software:book:blog}. In prior work, researchers have acknowledged the value of Internet artifacts in deriving practices, and analyzed Internet artifacts to summarize security practices used in DevOps~\cite{rahman:csed2017:devsecops}, practices used for continuous deployment~\cite{rahman:agile2015:cd}, and testing practices~\cite{garousi2018smells:blog}. Analysis of Internet artifacts can be useful for systematizing Kubernetes security knowledge---a research topic that remains under explored~\cite{brad:rsa:k8s:underexplored}. By systematically analyzing Internet artifacts related to Kubernetes security we hypothesize to derive a list of security practices.   


\textit{The goal of this paper is to help practitioners in securing their Kubernetes installations through a systematization of knowledge related to Kubernetes security practices.} 
 
We answer the following research question: \textbf{\textit{RQ:} What Kubernetes security practices are reported by practitioners?}

We systematize knowledge related to Kubernetes security by conducting a grey literature review~\cite{grey:original} where we apply qualitative analysis on Internet artifacts. We collect required Internet artifacts using the Google search engine with three search strings. Next, we apply a set of filtering criteria and apply qualitative analysis~\cite{saldana2015coding} on 104 Internet artifacts, such as blog posts, to construct a list of security practices. 

We list our contributions as following: 

\begin{itemize}[leftmargin=*]
    \item{A synthesized list of security practices for Kubernetes; and}
    \item{A curated dataset\cite{k8s:dataset} with a mapping between Internet artifact and identified security practices.}
\end{itemize}
 
The rest of the paper is organized as follows: in Section \ref{meth}, we provide methodology of our paper. In Section \ref{taxonomy-res}, we describe our derived Kubernetes security practices in details. In Section \ref{related}, we discuss prior research on Kubernetes. We discuss our findings with implication for users and researchers and conclude the paper in Section \ref{discussion}.



\begin{figure*}[h]
\centering 
\includegraphics[scale=0.95]{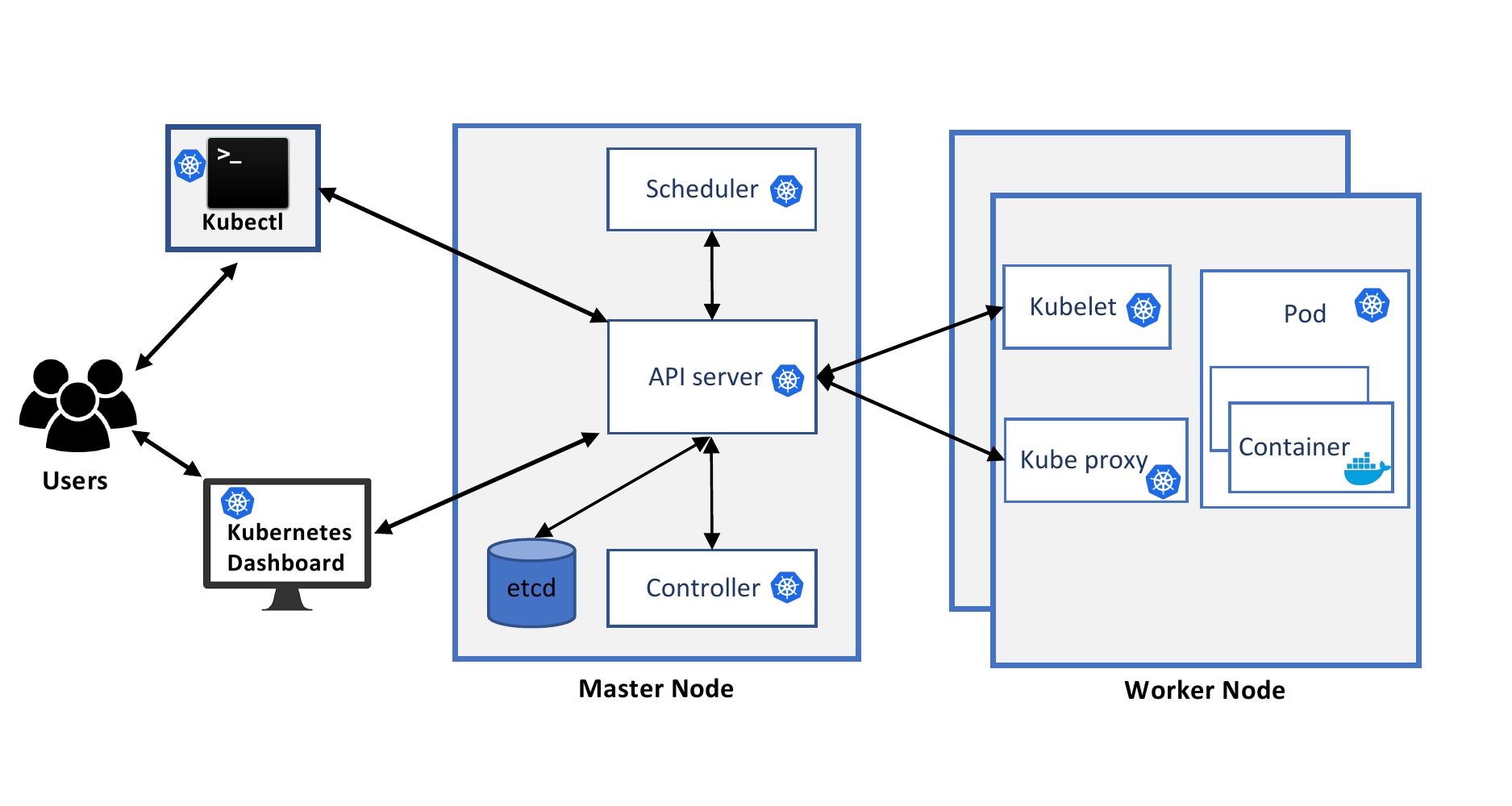}
\caption{A brief overview of Kubernetes. Kubernetes users interact with the installation using the Kubernetes dashboard and `kubectl'. The purpose of master node is to maintain the desired cluster state and manage worker nodes. Worker nodes are used to run containerized applications inside the pod.}
\label{fig-k8s}
\end{figure*}


\section{Methodology} 
\label{meth}


We first provide background on Kubernetes in Section~\ref{background}. Next, we provide methodology details in Section~\ref{taxonomy-meth}. 

\subsection{Background}
\label{background} 
 
Kubernetes is an open-source software for automating management of computerized services such as containers~\cite{miles:book2020:kubernetes}. A Kubernetes installation is colloquially referred to as a Kubernetes cluster~\cite{miles:book2020:kubernetes}. Each Kubernetes cluster contains a set of worker machines defined as nodes. As shown in Figure~\ref{fig-k8s}, two types of nodes exist for Kubernetes: master nodes and worker nodes.

Each master node includes the following components: `API server', `scheduler', `controller', and `etcd'~\cite{miles:book2020:kubernetes}. The `API server' is responsible for orchestrating all the operations within the cluster. Kubernetes serves its functionality through an application program interface from the `API server'. The `controller' is a component on the master that watches the state of the cluster through the `API server' and changes the current state towards the desired state. The `scheduler' is the component in the control plane responsible for scheduling pods across multiple nodes. The `etcd' is a key-value based database that stores all configuration information for the Kubernetes cluster. Users use a command-line tool `Kubectl' to communicate with the `API server' in the master node.

The worker nodes host the applications that run on Kubernetes~\cite{miles:book2020:kubernetes}. The following components are included in the worker node: `kube-proxy', `kubelet' and `pod'. `kube-proxy' maintains the network rules on nodes. `kubelet' is an agent that ensures containers are running inside a pod. The pod is the smallest Kubernetes entity, which includes at least one active container. A container is a standard software unit that packages the code and associated dependencies to run in any computing environment~\cite{miles:book2020:kubernetes}.

\subsection{Methodology to Identify Kubernetes Security Practices} 
\label{taxonomy-meth} 

We synthesize Kubernetes security practices by conducting a grey literature review~\cite{grey:original}. A grey literature review is the process of reviewing and synthesizing content included in Internet artifacts, such as blog posts and video presentations~\cite{grey:original}. A grey literature review is different from a systematic mapping study or systematic literature review, as in these types of literature reviews, researchers use peer-reviewed scientific articles indexed in scholar databases. In prior work, researchers have reported that practitioners use Internet artifacts, such as blog posts to report their experiences, recommendations, and the practices they follow. Previously, researchers have systematically studied Internet artifacts to identify challenges in microservices development, identify practices used in continuous deployment~\cite{rahman:agile2015:cd}, identify security practices used in organization who have adopted DevOps~\cite{rahman:csed2017:devsecops}, and software testing~\cite{GAROUSI:grey:testing}. Our hypothesis is that by systematically analyzing Internet artifacts we can synthesize Kubernetes security practices reported by practitioners.  


We conduct grey literature review using the following steps: 

\noindent \textbf{\textit{Step\#\MYROMAN{1}}-Collect Internet Artifacts}: We use the Google search engine to collect our Internet artifacts. We use 3 search strings: `kubernetes security practices', `kubernetes security good practices', and `kubernetes security best practices'. We start with the search string `kubernetes security practices', and later on added the other 2 search strings because while collecting search results with the first string we observe practices being referred to as `good practices' and `best practices'. After performing the search we collect the first 100 search results, as Google displays the results in a sorted order based on relevance.

\noindent \textbf{\textit{Step\#\MYROMAN{2}}-Select Internet Artifacts}: We apply an inclusion criteria on the collected search results to identify Internet artifacts that discuss security practices for Kubernetes. The inclusion criteria is listed below: 
\begin{itemize}[leftmargin=*]
\item{The Internet artifact is not a duplicate}; 
\item{The Internet artifact is available for reading}; and 
\item{The Internet artifact discusses security practices for Kubernetes};
\end{itemize}

\noindent \textbf{\textit{Step\#\MYROMAN{3}}-Qualitative Analysis}: We use open coding~\cite{saldana2015coding}, a qualitative analysis technique, to determine the security practices for Kubernetes. In open coding a rater observes and synthesizes patterns within unstructured text~\cite{saldana2015coding}. To determine the practices, the first author apply open coding on the content of the Internet artifacts to derive the security practices. The first author is a graduate student with a professional experience of one year in Kubernetes, and one year of academic experience in software security. 

\noindent \textbf{\textit{Step\#\MYROMAN{4}}-Verify Rating}: The process of determining the practices is susceptible to first author bias. We mitigate this bias by allocating another rater, the second author of the paper, who apply closed coding~\cite{crabtree:coding:book} on a randomly selected set of 50 Internet artifacts. Closed coding is the technique of mapping an entry to a pre-defined category~\cite{crabtree:coding:book}. For each of the 50 Internet artifacts, the second author examined if the artifact of interest includes a discussion related to the security practices identified by the first author. The second author has 3 years of experience in software security. We calculate the agreement rate between the first and second author for the 50 Internet artifacts using Cohen's Kappa~\cite{cohens:kappa}.


\section{Kubernetes Security Practices} 
\label{taxonomy-res} 

In this section we answer: \textbf{\textit{RQ: What Kubernetes security practices are reported by practitioners?}} After applying open coding on 104 Internet artifacts we derive 11 practices for Kubernetes security. Of the 104 Internet artifacts 90.38\%, 4.81\%, and 4.81\% are respectively blog posts, videos and presentations. We describe each of these practices below, where the count of Internet artifacts is enclosed within parenthesis:

\noindent \MYROMAN{1}. \textbf{Authentication and Authorization (82)}: The practice of applying authentication and authorization rules to prevent malicious users from getting access and performing unauthorized activities inside the Kubernetes cluster. Authentication in Kubernetes refers to the authentication of API requests through authentication plugins\cite{k8s:docs}. Authorization in Kubernetes refers to the evaluation of each authenticated API request against all policies to allow or deny the request\cite{k8s:docs}. Practitioners have reported a set of tasks to implement the practice of authentication and authorization:
    
    \begin{itemize}[leftmargin=*]
        \item{Anonymous access to the Kubernetes server needs to be disabled. By default, Kubernetes allows anonymous access to the Kubernetes API server. \cite{k8s:docs}}
        
        \item{Default authorization modes need to be disabled. }
        
        \item{Admission controllers need to be enabled. In Kubernetes, an admission controller is a tool that intercepts requests to the Kubernetes API after the request is authenticated and authorized, but before a volume is made persistent.}  
        
        \item{Controlling the use of impersonation: Kubernetes allows one user to act as another user through impersonation headers\cite{k8s:docs}. The impersonation feature has benefits, for example, a user designated as an admin can use this feature to debug authorization by impersonating another user and  checking if the request was denied. However, in case of failure to define limitations on who can impersonate and what the impersonated user can do, the impersonation feature can be detrimental to the security of Kubernetes.}
        
        \item{Default configurations must be changed. The use of default configuration in authentication and authorization can allow any anonymous unauthenticated user to perform malicious activities. For example, a malicious user can guess the default configuration of an insecure admission, gain access to the admission controller, and run malicious commands.} 
    \end{itemize}
    
     For authentication and authorization, practitioners suggest the use of OpenID~\footnote{https://openid.net/}, a standard protocol for authentication. The official Kubernetes documentation also provides guidelines on how to implement secure authorization using webhooks, role-based access control (RBAC) and attribute-based access control (ABAC)~\cite{k8s:docs}.
    

\noindent \MYROMAN{2}. \textbf{Implementing Kubernetes-specific Security Policies (81):} The practice of applying policies to secure Kubernetes components, pods and network of Kubernetes clusters to prevent security breaches. 
    \begin{itemize}[leftmargin=*]
        
        \item{\textit{Network-specific policies}: The practice of applying a network policy to protect communication between Kubernetes pods from undesirable network communications. By default, all Kubernetes pods can communicate with other pods. Practitioners recommend policies to restrict traffic between pods, restrict API server access and reducing network exposure to secure the network. If network policies are not defined and firewalls are not set, then anyone may attack the API server from any IP address. Practitioners also suggest imposing proper firewalls to block all undesirable network communication using network policy plugins like Calico ~\footnote{https://www.projectcalico.org/calico-networking-for-kubernetes/} and configuring restricted access to a database for pods.}
        
        \item{\textit{Pod-specific policies}: The practice of implementing a policy for pods to apply security context to pods and containers. Pod policies determine how the workloads should run in the Kubernetes cluster. Without defining a secure context for the pod, a container may run with root privilege and write permission into the root file system, which can make the Kubernetes cluster vulnerable. Practitioners recommend containers inside a pod must run as a non-root user with read-only permission and enabling Linux security modules. Practitioners also recommend that users install the minimal version of operating systems to reduce the attack surface.}

        \item{\textit{Generic policies}: The practice of applying a generic security policy to protect Kubernetes cluster components from external malicious users. TCP ports for kubelet, API server, etcd, and network plugins should not be left open and should require authentication to have visibility. Every user in the system should have the least privilege by default. Public SSH access to Kubernetes cluster nodes should be restricted. Practitioners recommend that Kubernetes users create an audit policy for logging, and audit policies must be configured for each Kubernetes cluster at the API server level.  }

    \end{itemize}

\noindent \MYROMAN{3}. \textbf{Vulnerability scanning (63):} The practice of scanning Kubernetes components and continuous delivery (CD) components for vulnerabilities. 
    \begin{itemize}[leftmargin=*]
        
        \item{Kubernetes components, such as containers can contain vulnerabilities and malicious malware. If vulnerabilities are present in a Kubernetes cluster, then the entire container orchestration system, and the provisioned applications, become susceptible to attacks. For example, in 2017, researchers found Docker images embedded with malicious malware. Practitioners  recommended scanning containers for vulnerabilities with tools,such as `Dockscan'~\footnote{https://github.com/kost/dockscan} and `CoreOS Clair'~\footnote{https://github.com/quay/clair}.} 
        
        \item{If images and deployment configurations within CD components are not inspected, then it can make the Kubernetes cluster vulnerable to malicious users. The malicious users can gain access at a later point when these images are deployed and may exploit the latent vulnerabilities in Kubernetes production environments. Practitioners recommend pulling images from a trusted private registry and checking for the vulnerability of code and images.} 
    \end{itemize}

\noindent \MYROMAN{4}. \textbf{Logging (47):} The practice of enabling and monitoring logs for the Kubernetes cluster. Practitioners recommend that logging should be enabled for (i) applications, (ii) the containers within each pod, and for (iii) Kubernetes clusters for system health checking.  Without enabling logging and monitoring, users may face difficulty troubleshooting unexpected consequences, such as attacks from malicious users and outages. To implement the practice of logging, practitioners propose the following practices:  
    \begin{itemize}[leftmargin=*]
        \item{ Logs must be monitored at a regular interval. }
        \item{Alerts must be set up for any drastic change in log metrics comparing to previous log records.}
      
    \end{itemize}

\noindent \MYROMAN{5}. \textbf{Namespace separation (36):} The practice of separating namespaces so that the resource of one namespace are not shared with another. A `namespace' in Kubernetes is a logically isolated virtual cluster within the same physical cluster.\cite{k8s:docs} Creation of separate namespaces enables resources to be isolated between namespaces. If a separate namespace is not created for a resource then the resource gets `default' namespace. Practitioners recommend that each team in a company should have a separate namespace for better manageability and running its development and production environments\footnote{https://cloud.google.com/blog/products/gcp/kubernetes-best-practices-organizing-with-namespaces}. If there is only a `default' namespace and no separate namespace for different teams then any malicious user can perform an attack on the `default' namespace making the entire resource vulnerable to that attack. Practitioners use the \textit{--namespace} flag in kubectl command to separate namepsaces.  

\noindent \MYROMAN{6}. \textbf{Encrypt and restrict access to etcd (34):} The practice of encrypting and restricting access to `etcd', the internal database used by Kubernetes\cite{k8s:docs}. Practitioners recommend `etcd' to only be available from the API servers, and to be isolated behind a firewall so that outsiders can not get access via API. 

By default, Kubernetes stores secret data as plaintext in `etcd'\footnote{https://ubuntu.com/kubernetes/docs/encryption-at-rest}. In that case, if a malicious user gets access to `etcd', then the malicious user can retrieve sensitive information, such as database user names, passwords, and queries. Although Kubernetes does encrypt `etcd', the key for the encryption is stored as plaintext in the config file in the master node. For that reason, practitioners recommend using secret management tools for additional security\cite{k8s:docs}, such as `Vault'\footnote{https://www.vaultproject.io} for encryption.

\noindent \MYROMAN{7}. \textbf{Continuous update (28):} The practice of applying security patches to keep the Kubernetes cluster updated with latest security fixes. Practitioners recommend that Kubernetes users apply updates as well as conducting continuous updates for the deployed applications within the Kubernetes pods. Without continuous updates, vulnerabilities might exist in the Kubernetes installation, which can give malicious users opportunity to perform attacks. 

Vulnerabilities in Kubernetes are not uncommon: for example, two vulnerabilities CVE-2019-16276~\cite{cve-2019-16276} and CVE-2019-11253~\cite{cve-2019-11253} were discovered in October 2019 in Kubernetes~\footnote{https://security.berkeley.edu/news/kubernetes-vulnerabilities-allow-authentication-bypass-dos-cve-2019-16276}. The vulnerability `CVE-2019-16276' was related to `CWE-444 Inconsistent Interpretation of HTTP Requests (`HTTP Request Smuggling')'. The vulnerability `CVE-2019-11253' was related to `CWE-20 Improper Input Validation'. The security patches for `CVE-2019-11253' and `CVE-2019-16276' were released on October 16, 2019 and October 22, 2019 respectively~\footnote{https://cloud.google.com/kubernetes-engine/docs/security-bulletins}. If any Kubernetes user does not install these security patches then the Kubernetes cluster will be susceptible to a denial of service attack. 

For continuous updates, practitioners have also recommended the use of rolling update, i.e. installing Kubernetes patches without disrupting the availability of the deployed applications.\footnote{https://k8s.vmware.com/kubernetes-security-best-practices/} Kubernetes provides tools, such as `kubectl' to perform rolling updates~\cite{k8s:docs}. 

\noindent \MYROMAN{8}. \textbf{Limit CPU and memory quota (18):} The practice of limiting CPU and memory to a pod or a namespace so that malicious attacks can be mitigated. By default, all resources in Kubernetes start with unbounded memory requests/limits and unbounded CPU access. If a malicious user starts a denial of service (DOS) attack with in a pod within the Kubernetes cluster then, due to a high volume of requests, kube-scheduler will create a new pod and an instance of the container will start inside the new pod. This process continue until it consumes all available CPU resources and memory leaving all the applications in starvation. Hence, failure to define CPU and memory request limits for a pod or the namespace may result in a consumption of all available resources in the Kubernetes cluster, enabling a denial of service (DOS) attack. 

Practitioners can configure the amount of resources by defining a maximum number of instances for a  container, the number of CPU share for an application to consume, and the maximum amount of memory for a pod or namespace.

\noindent \MYROMAN{9}. \textbf{Enable SSL/TLS support (18):} The practice of enabling secure sockets layer (SSL) or transport layer security (TLS) protocol to ensure secure and encrypted communication between Kubernetes components. Enabling TLS between kubernetes api server, etcd, kubelet and kubectl ensures secure communication between cluster components. Practitioners suggest enabling TLS and SSL certificates for Kubernetes components.
    
\noindent \MYROMAN{10}. \textbf{Separate sensitive workload (14):}  The practice of running sensitive applications on a dedicated set of machines to limit the potential impact of a security breach. For example, if a malicious user gets access to a node's `kubelet' credentials, then the user can access the contents of secrets and gain control of the entire file system, but the user will not be able to access the sensitive applications and associated secrets. Practitioners recommend Kubernetes-provided utilities, such as `taints and tolerations'\cite{k8s:docs} that can control where a pod might be deployed. 

\noindent \MYROMAN{11}. \textbf{Secure metadata access (9):} The practice of securing the sensitive metadata of the Kubernetes cluster. Practitioners state that the Kubernetes metadata APIs provide a gateway to expose `kubelet' admin credentials. Google recommends activating features such as  `Workload Identity’\footnote{https://cloud.google.com/kubernetes-engine/docs/how-to/protecting-cluster-metadata} for Google Kubernetes Engine (GKE) to prevent any sensitive information from leaking through the metadata service.

    
\textit{\textbf{Rater verification}}: The Cohen's Kappa between the two raters is 0.8, which is substantial according to Landis and Koch~\cite{Landis:Koch:Kappa:Range}.


\section{Threats to Validity}
\label{threats} 

We discuss the limitations of our paper as following: 

\noindent \textbf{Conclusion Validity}: Our derived set of practices is limited to our collection of 104 Internet artifacts. Our collection of Internet artifacts might have missed Internet artifacts, that may have included practices not identified in our paper.  We mitigate this limitation by systematically collecting a set of 104 Internet artifacts.  

The identified practices are also susceptible to biases of the rater who identified the practices by applying open coding. We mitigate this limitation by allocating another rater, who applied closed coding. The Cohen's Kappa between the two raters is 0.8. which is substantial~\cite{Landis:Koch:Kappa:Range}.

\noindent \textbf{Construct Validity}: Our identified categories are susceptible to experimenter bias. The first author who derived the practices has professional experience in Kubernetes. The first author's professional experience can formulate expectations related to security practices for Kubernetes, which may influence the identified practices.   
    
\noindent \textbf{External Validity}: Our findings might not be generalizable as we might have excluded practices unique to the proprietary domains, and not discussed publicly in Internet artifacts.


\section{Related Work}
\label{related} 

Our paper is related to prior research that has investigated usage and maintenance of Kubernetes. Burns et al.~\cite{k8s:original:google} described the evolution of container management systems at Google, and described how two initial internal systems called Borg and Omega was evolved into Kubernetes. Brewer~\cite{k8s:brewer} conducted a case study on Kubernetes and discussed how key concepts of Kubernetes can be used to simplify scaling of containers. Medel et al.~\cite{medel:k8s:performance} used real data collected from Kubernetes and applied formal modeling to characterize performance and resource management in Kubernetes. Chang et al.~\cite{chang:k8s:monitoring } constructed a monitoring platform to dynamically provision cloud resources using Kubernetes. Vayghan et al.~\cite{vayghan:k8s:outage} investigated availability of Kubernetes using a set of experiments, and reported that service outages can occur frequently. Shah and Dubaria~\cite{shah:k8s:compare} compared orchestration management features of Docker Swarm, Kubernetes, and Google Cloud Platform, and observed Kubernetes to provide features, such as deployment, monitoring, and easy scalability. Takahashi et al.~\cite{takahashi:k8s:portable} proposed a portable load balancer for Kubernetes, and reported improved portability without sacrificing performance. Song et al.~\cite{song:k8s:api} used Kubernetes to construct an auto scaling system for API gateways. The authors~\cite{song:k8s:api} report that their constructed system improves utilization of system resources, while ensuring high availability. Muralidharan et al.~\cite{murali:k8s:iot} constructed a Kubernetes-based system to monitor and manage Internet of Things (IoT) applications for smart cities. Wei-guo et al.~\cite{wei:k8s:scheduling} constructed a resource scheduling algorithm for Kubernetes using ant colony and particle swarm optimization techniques. The scheduling algorithm proposed by Wei-guo et al.~\cite{wei:k8s:scheduling} outperforms the original algorithm used in Kubernetes.   

The above-mentioned discussion highlights Kubernetes research in two areas: (i) use of Kubernetes in creating systems, such as monitoring systems and (ii) case studies on Kubernetes related to performance and resource management. We observe a lack of research related to security practices for Kubernetes. We address this research gap by systematically synthesizing practitioner-reported security practices using grey literature review.


\section{Discussion and Conclusion}
\label{discussion} 

\noindent \textbf{With great power comes responsibilities}: Kubernetes provide utilities for users to manage containers at scale. However, our description of the 11 practices in Section~\ref{taxonomy-res} shows that effective and secure usage of Kubernetes requires the implementation of security practices applicable for multiple components within the Kubernetes installations: containers, pods, `etcd' database etc. The application of the aforementioned 11 practices also need a deep understanding of Kubernetes components and configurations. Our discussion in Section~\ref{taxonomy-res} can be helpful in two ways: \textit{first}, understand the components where security practices are applicable. \textit{Second}, practitioners who already have Kubernetes in place, can use our identified practices as a benchmark and compare their usage of practices.     

\noindent \textbf{Implication for researchers}: Our discussion in Section~\ref{related} shows that Kubernetes security to be an under explored research area. Our derived list of security practices can provide the groundwork for future research in Kubernetes security. To what extent the reported security practices are in use can be quantified systematically. Researchers can measure the attack surface associated with Kubernetes components and configurations. Researchers can find possible mitigation strategies such as static analysis and dynamic analysis to inspect insecure practices in Kubernetes. Researchers can also quantify how frequently the identified practices are actually in use. Furthermore, detection and mitigation of security misconfigurations that occur in Kubernetes could be of interest to researchers.     

\noindent \textbf{Conclusion}\label{conclusion}: As Kubernetes usage becomes increasingly popular, securing Kubernetes is of paramount importance to practitioners. A systematization of knowledge related to practitioner-reported practices might be helpful to secure Kubernetes installations. We conduct a qualitative analysis of Internet artifacts, such as blog posts to identify 11 security practices for Kubernetes. Our derived list of practices include continuous update, enable SSL/TLS support, vulnerability scanning and logging. Our paper can help practitioners in securing Kubernetes installations. Further, our findings can lay the groundwork to conduct research in Kubernetes security.  



\balance 

\bibliographystyle{IEEEtran}
\bibliography{k8s}

\end{document}